\documentclass[12pt,preprint]{aastex}

\shorttitle{HI in the PG1216+069 sub-DLy$\alpha$ absorber
$z=0.00632$}
\shortauthors{Briggs and Barnes}

\begin{document}

\title{HI 21cm observations of the PG1216+069 sub-DLy$\alpha$ absorber field at
$z=0.00632$}

\author{F. H. Briggs}
\affil{Research School of Astronomy and Astrophysics, The Australian National University, Mount Stromlo 
Observatory, Weston Creek, ACT 2611, Australia \and
Australian Telescope National Facility, CSIRO, Epping, NSW, Australia}
\email{fbriggs@mso.anu.edu.au}
\and
\author{D.G. Barnes}
\affil{Department of Physics, University of Melbourne, Melbourn, VIC 3010, Australia}
\email{dbarnes@physics.unimelb.edu.au}

%...............................................................

\begin{abstract}
The Westerbork Synthesis Radio Telescope finds a weak 21cm line emission feature at the
coordinates (RA-Dec-velocity) of the
sub-Damped Lyman-$\alpha$ absorber observed at $z_{abs}=0.00632$ in the spectrum of
PG1216+069. The emission feature, WSRT-J121921+0639, lies within $30''$ of the quasar sightline,
is detected at 99.8\% (3$\sigma$) confidence level,
has $M_{HI}$ between 5 and 15${\times}10^{6}$M$_{\odot}$, and has velocity spread
between 20 and 60~km~s$^{-1}$. Other HI emitters in the field include VCC297
at a projected distance of $86h^{-1}_{75}$kpc
% with $M_{HI}\sim 1.5{\times}10^{8}$M$_{\odot}$
and a previously unreported HI cloud , WSRT-J121919+0624 at $112h^{-1}_{75}$kpc 
%($XXh^{-1}_{75}$kpc projected distance  from PG1216+069)
with $M_{HI}\sim 3{\times}10^{8}$M$_{\odot}$. The optically
identified, foreground galaxy that is closest to the quasar sightline
appears to be VCC339 (${\sim}L^*/25$) at $29h^{-1}_{75}$kpc and velocity offset 292~km~s$^{-1}$ .
A low surface brightness galaxy with the HI mass of the sub-DLA absorber
WSRT-J121921+0639 would likely have
m$_B\sim 17$, and its  diffuse optical emission would need to compete with the light of both
the background QSO and a brighter foreground star $\sim$10$''$ from the QSO sight line. 
\end{abstract}


%...............................................................

\section{Introduction}

The goal of clarifying the nature of QSO absorption line systems 
drives the interest in extreme low redshift 
absorption systems  such as the ``sub-Damped Lyman-$\alpha$'' absorber in the sight line to the
$z=0.33$ AGN PG1216+069. At a heliocentric redshift velocity of 1895~km~s$^{-1}$, the absorber falls in the
outskirts of the Virgo Cluster, where its association with a particular galaxy host or group/cluster environment should
become clear through observation.  Tripp et al.s'  recent, comprehensive study of PG1216+069 \citep{tripp05} 
makes it clear that 
(1) the metallicity  is low, [O/H]$\approx$-1.6, but at a  characteristic level for  high redshift absorbers, 
(2) there is no detected optical emission from the absorber itself  nor an obvious association with a  specific  galaxy in the field, and 
(3) the HI column density through the absorber $N_{HI} \sim 10^{19.3}$cm$^{-2}$ is a factor 10 below the Damped Lyman-$\alpha$ (DLA) class. 
While the lower limit to DLA column density $10^{20.3}$cm$^{-2}$
was originally set as a completeness limit in the early surveys \citep{wolfe86}, this column density also
takes on physical significance in disk galaxies, since it is typical of the threshold column density for the onset
of star formation \citep{kennicutt89}. \citet{zwaan05} used an extensive data set of 21cm line observations to
demonstrate the consistence between QSO absorption line statistics and the cross sections that $z\approx 0$
galaxies present for absorption. Zwaan et al. further demonstrate that the decline in metal abundance for 
low column density absorbers agrees with abundance gradients in the outskirts of nearby galaxies.

QSO absorption statistics indicate through the $f(N)$ distribution \citep{tytler87,petitjean93} that populations of low column density
absorbers present larger cross sections than the DLA class, implying that clouds with $N_{HI} \sim 10^{19}$cm$^{-2}$
may be more common than the higher $N_{HI}$ ones in spiral galaxy disks.
Radio telescopes  routinely 
observe $N_{HI}$ well below this level, thereby providing the
incentive for a WSRT reconnaissance into the PG1216+069 field.  \citet{kanekar05} observed this field  with GMRT within the same time frame without detecting the new features we report here.
 
\section{The WSRT Observations}

The Westerbork Synthesis Radio Telescope (WSRT) observed the field centered on
PG1216+069 in  ``filler time''  scheduled through the interim proposal system.
Rather than a single long integration in one 12 hour integration, the observation consists
of two shorter runs. On  2004 August 14,  the first session, covering extreme negative hour angles, produced 4.5 hours on source in the ``2${\times}$144m''
configuration (with RT9 to movable telescope distances of 144, 288, 1368, and 1440 meters).
On 2004 August 27, 
the second session covered 4 hours of extreme positive hour angles in the ``Maxi-short'' configuration
(RT9 to movable telescopes distances of  36, 90 133, 1404 meters). The DZB spectrometer 
analyzed both linear polarizations with a 10 MHz ($\sim$2000 km~s$^{-1}$) band centered on heliocentric redshift velocity 1900~km~s$^{-1}$ with 256 spectral channels per polarization to obtain channel
spacing of 8.3~km~s$^{-1}$ or velocity resolution of 16.5~km~s$^{-1}$ after Hanning smoothing. 

For an east-west array such as WSRT, observing sources close to the celestial equator has
two  important consequences: The negative one is that the telescope resolution in the north-south
direction deteriorates
at low declination $\delta$  as $\csc\delta$, so that the best resolution possible for PG1216+069
is a north-south fan beam of 108$''\times$13$''$. The positive consequence is that all projected baselines foreshorten
at large hour angle $h$ as $\cos h$, so that the apparent lack of short spacings in the  ``2${\times}$144m'' configuration data does not damage the surface brightness sensitivity to extended emission.
Short spacing sensitivity is limited to  $>$25m projected spacings in order to avoid
shadowing of one dish by its neighbor, causing the observation to be 
insensitive to smooth structure on angular scales larger than $\sim$15 arcminutes (115~kpc).
 
Data processing steps included editing, calibration and application of different weighting schemes in the synthesis imaging software packages AIPS and
MIRIAD.  This produced spectral line data cubes with a range of  resolution and noise level,
summarized in Table~\ref{noiselevel.tab}.  For each of three weighting schemes, the table lists the
angular resolution and noise level $\sigma_S$ measured from the processed data cubes. Assuming that
the emission fills the synthesized beam solid angle $\Omega_B$  leads  to 
a surface brightness sensitivity ($\Delta T_B = \sigma_S \lambda^2/2{\rm k_B}\Omega_B$) and
column density sensitivity ($\Delta N_{HI} = 1.8{\times}10^{18}\Delta V_{\rm km/s} \Delta T_B$ cm$^{-2}$).
The table includes estimates of the flux density that a beam filled with $N_{HI}\sim 10^{19.3}$cm$^{-2}$
would produce, along with the HI mass contained within the beam 
($M_{HI} = 2.3{\times}10^5\Delta V_{\rm km/s} S_{\rm Jy} d^2_{\rm Mpc}$M$_{\odot}$), assuming
the distance of 26.6~Mpc, which \citet{tripp05} adopted.

Analysis of the continuum image formed from the average of the channels free of HI signal found 
that PG1216+069 had a continuum flux density of 2.7$\pm$0.2~mJy at 1416 MHz (consistent with
the 2.8$\pm $0.3~mJy  measurement of \citet{kanekar05}). The NVSS survey
\citep{condon98}
indicates an upper limit $\sim$1.9~mJy (${\sim}2\sigma$), while VLBA observations indicated compact structure
of 5.7$\pm$0.3~mJy in the period 1999-2000 \citep{ulvestad05}. The PG1216+069 AGN appears to be a variable radio 
source.

\begin{figure}
\epsscale{1.2}\hglue -1.3cm
\plotone{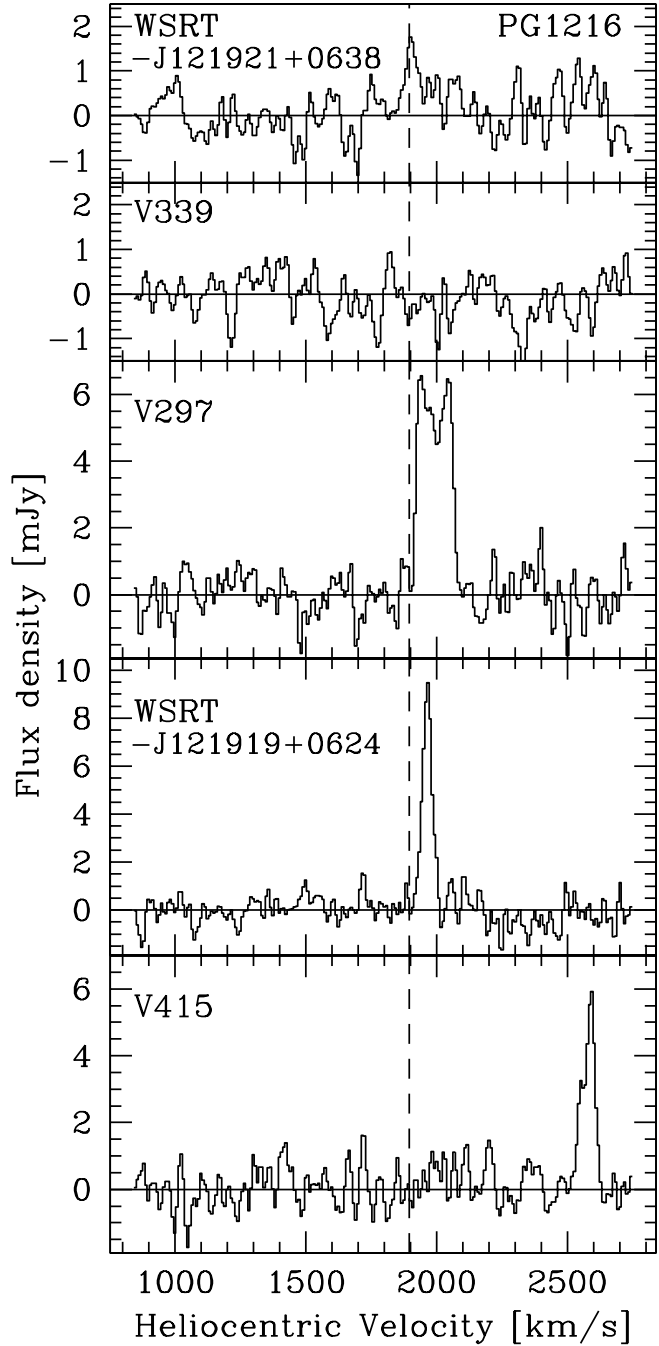}
\caption{HI spectra for selected objects in the PG1216+069 field. Vertical dashed line
indicates the absorption line redshift. Spectra for WSRT-J121921+0638 and VCC339 have
been smoothed with a gaussian kernel with FWHM of 2 channels (16.6 km~s$^{-1}$); 
other spectra are unsmoothed.
\label{spectra.fig}}
\end{figure}

Figures~\ref{hi_field.fig} and  \ref{spectra.fig} summarize the observational results. 
Fig~\ref{hi_field.fig} presents contours of the integral line emission, obtained from the zeroth moment
of the spectral channels after automated masking using a smoothed version of the cube.
The image is 1 degree on a side, corresponding to 460~kpc. The half power beam width 
of the WSRT primary beam is $\sim$35~arcmin, so the outer regions of HI map  include the
first null of the pattern. Table~\ref{galaxies.tab} lists the galaxies with redshift less than
3000~km~s$^{-1}$ within 25 arcmin of the PG1216+069 sight line; many of these are
marked in Fig~\ref{hi_field.fig}.
Figure~\ref{spectra.fig} presents spectra for selected areas of the image.

\section{Discussion}

Figure~\ref{hi_field.fig} shows three significant detections for systems VCC297, VCC415, and
a previously unreported HI emitter, here named WSRT-J121918+0624,  centered at coordinates
12$^h$19$^m$17$^s$.6, 06$^{\circ}$24$'$06$''$ (J2000), close in the sky to VCC327.
The HIPASS online database\footnote{http://www.atnf.csiro.au/research/multibeam/release/}
shows  this feature %at $\sim$1975~km~s$^{-1}$ 
with $5\sigma$ significance. The HI mass in WSRT-J121918+0624 amounts to 
${\sim}3{\times}10^8M_{\odot}$ contained in a gaussian profile of FWHM $\sim$38~km~s$^{-1}$
centered on $V_{hel}=$ 1963~km~s$^{-1}$,
but the 800~km~s$^{-1}$ difference in redshift from VCC327 suggest that this is but chance
alignment between two unassociated objects. However, the 
NED\footnote{http://ned.ipac.caltech.edu/} indicates that there
 are at least  seven objects
(NGC4223, VCC287, VCC297, VCC332, WSRT-J121918+0624, VCC339, VCC346)
with velocities in the 400~km~s$^{-1}$ range
1870 to 2270 within a 300~kpc square area projected on the sky.  There is weak evidence 
 that WSRT-J121918+0624 is extended toward the
 NE, in which case the morphology would be typical of tidal debris from interactions within the 
 overdensity, similar to the structures seen in Ursa Major \citep{verheijen05} and elsewhere in Virgo
  \citep{chengalur95,oosterloo05}.
 
 The catalogued galaxy closest to the PG1216 sight line is VCC329 with a velocity offset
 of  292 ~km~s$^{-1}$ relative to the sub-DLA, but there is no  hint of an HI signal in
 the WSRT data (see Fig~\ref{spectra.fig}).
 
 Inspection of the spectrum along the line of sight to PG1216+069 reveals a weak emission
 feature at $V_{hel}=1905$, overlapping the velocity of the sub-DLA absorber. 
 The peak strength is 1.7$\pm$0.55mJy;
 for well behaved noise, there is only a 0.14\% probability of a 3$\sigma$ positive deflection
 occurring in a pre-chosen pixel by chance. The velocity width is not well constrained, and
 considering  widths in the range 20-60 km~s$^{-1}$ would indicate HI masses
 of 5 to 15${\times}10^6 M_{\odot}$. The strength of the feature is between 1.5 and 1.7~mJy per
 beam for all three synthesized beams in Table~\ref{noiselevel.tab}, implying that 
 a large fraction of the emission
 is likely to be contained within the $112''{\times}16''$ beam. Unless the HI spatial distribution
 is similar in shape to the elongated beam, the HI column density must vary in places
 to values above that specified by the damping wing profile of \citet{tripp05}.  
The signal strength measured by WSRT indicating 5${\times}10^6 M_{\odot}$ for a 16~km~s$^{-1}$ velocity width is compatible with the GMRT upper limit of  11${\times}10^6 M_{\odot}$ for a 20~km~s$^{-1}$ velocity width and $39.4''{\times}37.7''$ angular resolution.

 A rough rule of thumb for estimation of the optical brightness of a late type, gas rich galaxy is
 $M_B= -20.3 -2.8\log(M_{HI}/10^{10}M_{\odot})$ \citep{briggs90}. An $M_{HI}\approx 10^7M_{\odot}$ translates to $M_B\approx $-14.9, with large scatter.  With a distance modulus  equal 32.1, the  
integral optical emission would be
magnitude 17 and possibly much dimmer as would be typical of a low mass, diffuse LSB system. The optical emission must compete with the QSO  light (magnitude 15.84, 15.65,
and 15.61 in the B, $g$ and $i$ bands, respectively) and a star 10$''$ away that 
is tabulated at $g=14.9$ (according to \citet{kirhakos94} derived from an image in which it is saturated). 
The APMCat\footnote{http://www.ast.cam.ac.uk/$\sim$apmcat/} evaluates the sum of the star plus
QSO as $R=13.7$ and $B=14.9$.

\section{Conclusions}

\begin{figure}
\plotone{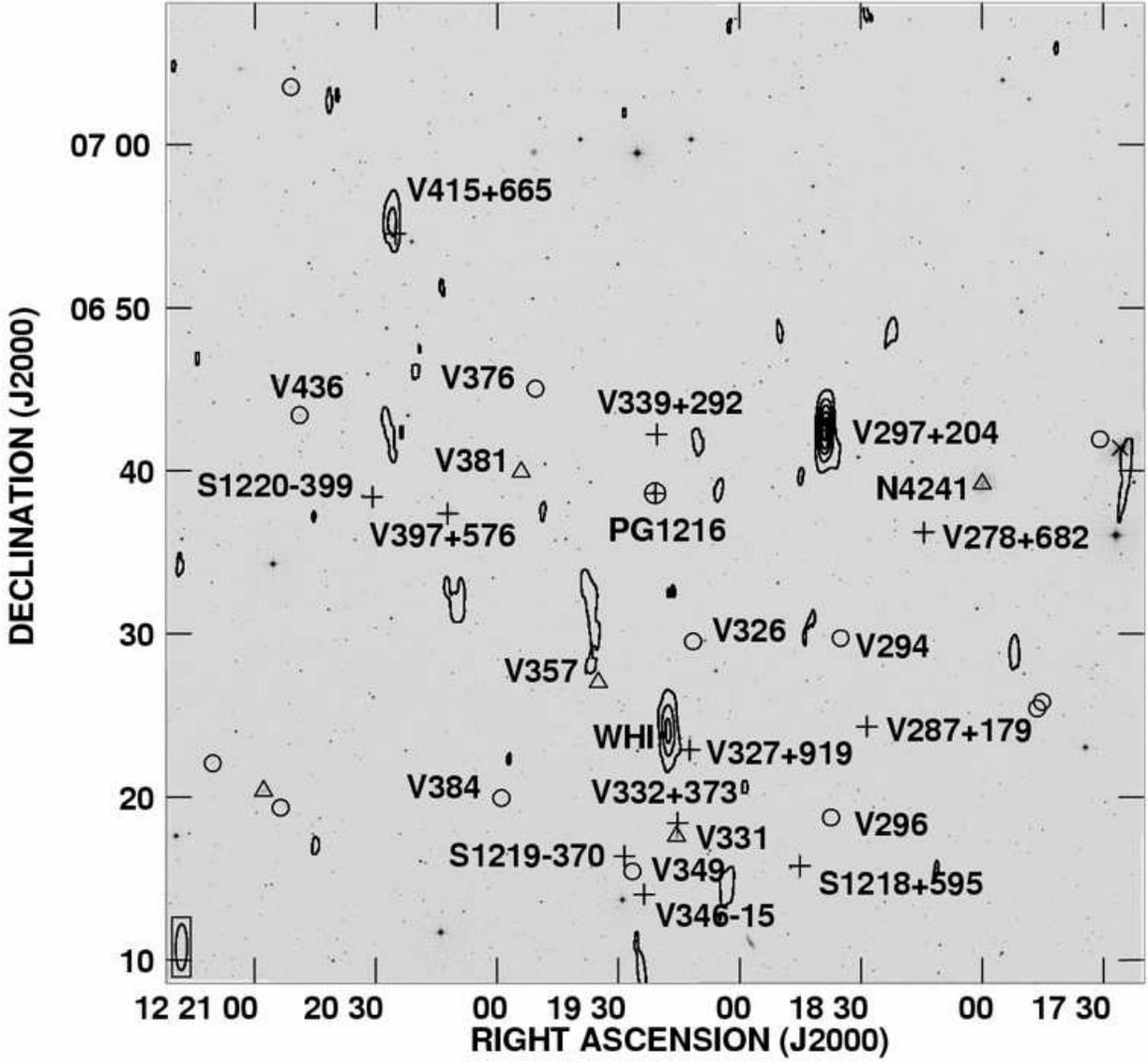}
\caption{Image of the PG1216+069 field in HI 21cm emission over the
heliocentric velocity range 900 to 2700 km~s$^{-1}$ with synthesized beam $169''{\times}42''$.
Contours of beam-averaged, integrated HI brightness overlay the grayscale
DSS optical image. Contours at 150, 300, 450, 600, 750 mJy~km~s$^{-1}$ per beam.
PG1216+069 falls at the center.
Labeled crosses indicate the galaxy centers for galaxies with redshifts between 895 
and 2895 km~s$^{-1}$, where the first part of the label is a galaxy ID  (see Table~\ref{galaxies.tab})
and the $\pm$XXX
indicates velocity relative to $V_{hel}=1895$~km~s$^{-1}$.  WHI indicates WSRT-J121918+0624
with velocity $+$68 relative to the sub-DLA redshift.
%VCC297 ($V_{hel}=1999$), VCC339 ($V_{hel}$ unknown but large), 
%VCC327 ($V_{hel}=2834$),
%VCC331 ($V_{hel}=642$), 
%VCC332 ($V_{hel}=2268$), 
%VCC415 ($V_{hel}=2560$),
%and
%PG1216+069 (absorption line $V_{hel}=1890$km~s$^{-1}$). 
Triangles mark VCC galaxies having redshifts
differing by more than 1000~km~s$^{-1}$ from $V_{hel}=1895$, while
circles mark VCC galaxies of unknown redshift. X indicates NGC4223
($V_{hel}=$ 2235),  far outside the primary beam.
Galaxies with $V_{hel}> 3000$km~s$^{-1}$ are omitted.
%LEVS= 3.0000E+01 * 2.000, 3.000, 4.000, 6.000, 9.000, 12.00
%$N_{HI} =$ 3.6, 4.7, 5.7, 7.2${\times}10^{18}$cm$^{-2}$.
%Peak in the image is 7.2722E+02 
\label{hi_field.fig}}
\end{figure}

The HI emission properties of the PG1216+069 sub-DLA absorber are consistent with expectations
based on the low column density, low metallicity and absence of optical emission from a clear host galaxy.  Low column density below the star formation threshold means that the lack of star formation locally along the line of sight is no surprise.  A low metallicity cloud with no connection to a optically identified host galaxy is compatible with the sub-DLA originating in a tiny, extreme low surface brightness dwarf of primitive composition or being a diffuse cloud in the outer halo of a galaxy such as VCC339 or VCC297.  The additional presence in this field
of the more massive cloud WSRT-J121918+0624 may argue for an origin for both features as tidal debris in this overdense volume. (The tranverse dimension of the field in Fig.~\ref{hi_field.fig} is 460~kpc, thus fitting easily within  Local Group, for example.)

Further progress
might be made by (1) longer integrations on this field with the goal of tracing low column density structures that may form a more obvious link to the visible galaxies in this overdensity and (2) more extensive, deep HI mapping in several other fields to identify examples of objects (dwarfs or halo clouds)
in more favorable situations (i.e., without optically bright QSOs cluttering the field).  Such populations
might account for the sub-DLA and Lyman-limit absorption cross sections.

\acknowledgments

We are grateful to the staff of the Westerbork Synthesis Radio Telescope, especially Rene 
Vermeulen and Tony Foley for coordinating the observations through the Interim Proposal
program.  The project was supported in part by grant DP0345001 from the Australian Research Council. This research has made use of the NASA/IPAC Extragalactic Database (NED) which is operated by the Jet Propulsion Laboratory, California Institute of Technology, under contract with the National Aeronautics and Space Administration. This research has made use of NASA's Astrophysics Data System. The Parkes telescope is part of the Australia Telescope which is funded by the Commonwealth of Australia for operation as a National Facility managed by CSIRO.

%Facilities: \facility{Nickel}, \facility{HST(STIS)}, \facility{CXO(ASIS)}.

\clearpage

\begin{table}
\begin{center}
\caption{Resolution and Sensitivity \label{noiselevel.tab}}
\begin{tabular}{cccccc}
\hline
$\theta_1{\times}\theta_2$ & $\Delta S$ & $\Delta T_B$ & $\Delta N_{HI}$ &
$S$(pred) & $M_{HI}$(pred)\\ 
arscsec$^2$ & mJy/beam & K  &  10$^{18}$cm$^{-2}$  & mJy& 10$^6$ M$_{\odot}$ \\
\hline
\hline
112$\times$16 & 0.57 & 0.27 & 8.2 &1.4 & 4\\
169$\times$42 & 0.61 & 0.07 & 2.2 &5.7& 15\\
235$\times$45 & 0.71 & 0.06& 1.8 &8.6 & 23\\
\hline 
\end{tabular} 
\end{center}
Notes: Columns give (1) half-power beam widths of synthesized beam, (2) rms noise level achieved in flux density in a single spectral channel, (3) rms sensitivity in brightness temperature, 
(4) rms sensitivity in HI columns density per 16.6~km~s$^{-1}$ velocity resolution, 
(5) the expected flux density if the beam were filled with $N_{HI}=10^{19.3}$cm$^{-2}$ emission
in 16.6~km~s$^{-1}$ ,
and (6) the HI mass contained in the beam, if filled with $N_{HI}=10^{19.3}$cm$^{-2}$.
\end{table}

\clearpage
\begin{table}
\begin{center}
\caption{Nearby Galaxies \label{galaxies.tab}}
\begin{tabular}{lcc}
\hline
Galaxy& Rel. Velocity & Proj. Separation\\ 
& km~s$^{-1}$ &  kpc \\
\hline
\hline
V339 &      292 &       29 \\
V381 &    -1414 &       65 \\
V326 &    ... &       72 \\
V376 &    ... &       76 \\
V297 &      104 &       86 \\
V357 &     1113 &       93 \\
V397 &      576 &       99 \\
V294 &    ... &      112 \\
W1216+064& 68 & 112\\
V327 &      919 &      123 \\
V278 &      682 &      129 \\
S1220+0638 &     -399 &      134 \\
V287 &      179 &      149 \\
NGC4241 &    -1162 &      155 \\
V332 &      373 &      156 \\
V384 &    ... &      161 \\
V331 &    -1238 &      162 \\
S1219+0616 &     -370 &      172 \\
V436 &    ... &      173 \\
V415 &      665 &      173 \\
V296 &    ... &      174 \\
V349 &    ... &      179 \\
S1218+0615 &      595 &      189 \\
V346 &      -15 &      190\\
\hline 
\end{tabular} 
\end{center}
Notes: Galaxy names use V to indicate the VCC, S to signify an abbreviated SDSS entry and W for WSRT.
The galaxy velocites are relative to the sub-DLA at 1895~km~s$^{-1}$. Projected separation of the sight line
assumes a distance of 26.6~Mpc.
\end{table}

\clearpage

\clearpage

\begin{thebibliography}{}
\bibitem[Briggs(1990)]{briggs90} Briggs, F.~H.\ 1990, \aj, 100, 
999 
\bibitem[Chengalur et al.(1995)]{chengalur95} Chengalur, J.~N., 
Giovanelli, R., \& Haynes, M.~P.\ 1995, \aj, 109, 2415 
 \bibitem[Condon et al.(1998)]{condon98} Condon, J.~J., Cotton, 
W.~D., Greisen, E.~W., Yin, Q.~F., Perley, R.~A., Taylor, G.~B., \& 
Broderick, J.~J.\ 1998, \aj, 115, 1693 
\bibitem[Kanekar \& Chengalur(2005)]{kanekar05} Kanekar, N., \& 
Chengalur, J.~N.\ 2005, \aap, 429, L51 
\bibitem[Kirhakos et al.(1994)]{kirhakos94} Kirhakos, S., Sargent, 
W.~L.~W., Schneider, D.~P., Bahcall, J.~N., Jannuzi, B.~T., Maoz, D., \& 
Small, T.~A.\ 1994, \pasp, 106, 646 
\bibitem[Kennicutt(1989)]{kennicutt89} Kennicutt, R.~C.\ 1989, 
\apj, 344, 685 
\bibitem[Oosterloo \& van Gorkom(2005)]{oosterloo05} Oosterloo, T., 
\& van Gorkom, J.\ 2005, \aap, 437, L19 
\bibitem[Petitjean et al.(1993)]{petitjean93} Petitjean, P., Webb, 
J.~K., Rauch, M., Carswell, R.~F., \& Lanzetta, K.\ 1993, \mnras, 262, 499 
\bibitem[Tripp et al.(2005)]{tripp05} Tripp, T.M., Jenkins, E.B., Bowen, D.V., Prochaska, J.X.,
Aracil, B., and Ganguly, R. 2005, \apj, 619, 714
\bibitem[Tytler(1987)]{tytler87} Tytler, D.\ 1987, \apj, 321, 49 
 \bibitem[Ulvestad et al.(2005)]{ulvestad05} Ulvestad, J.~S., 
Antonucci, R.~R.~J., \& Barvainis, R.\ 2005, \apj, 621, 123 
\bibitem[Verheijen \& Sancisi(2001)]{verheijen05} Verheijen, 
M.~A.~W., \& Sancisi, R.\ 2001, \aap, 370, 765 
\bibitem[Wolfe et al.(1986)]{wolfe86} Wolfe, A.~M., Turnshek, 
D.~A., Smith, H.~E., \& Cohen, R.~D.\ 1986, \apjs, 61, 249 
\bibitem[Zwaan et al.(2005)]{zwaan05} Zwaan, M.~A., van der 
Hulst, J.~M., Briggs, F.~H., Verheijen, M.~A.~W., \& Ryan-Weber, E.~V.\ 
2005, \mnras, 364, 1467 
\end{thebibliography}
\end{document}